\documentclass[preprint,showkeys, 12pt,amsmath,amssymb, aps]{revtex4-1}
\usepackage{amsthm}
\usepackage{bm}
\usepackage{graphicx}
\usepackage{multirow}
\usepackage{epstopdf}

\begin{document}

\title{Spheroidal and Nanocrystal Structures From Carbodiimide Crosslinking Reaction With RADA16}

\author{Jorge Monreal}
\email{jmonreal@alum.mit.edu}
\author{Robert Hyde}
\affiliation{Department of Physics, University of South Florida, Tampa, Florida, USA}

\begin{abstract}
RADA16 is a widely studied polypeptide known for its ability to self-assemble into $\beta$-sheets that form nanofibers.  Here we show that it is possible to self-crosslink the molecule via 1-ethyl-3-(3-dimethylaminopropyl)carbodiimide hydrochloride (EDC) as aqueous solutions.  The product results in a mix of nanocrystals and near micron-size spherules.  SEM and TEM pictures provide a view of the structures and nano tracking analysis give their size distributions.  FTIR analysis provides evidence for the existence of a crosslinking reaction.  
\begin{figure}[ht!]
\graphicspath{ {./DisPics/} }
\includegraphics [scale=0.5]{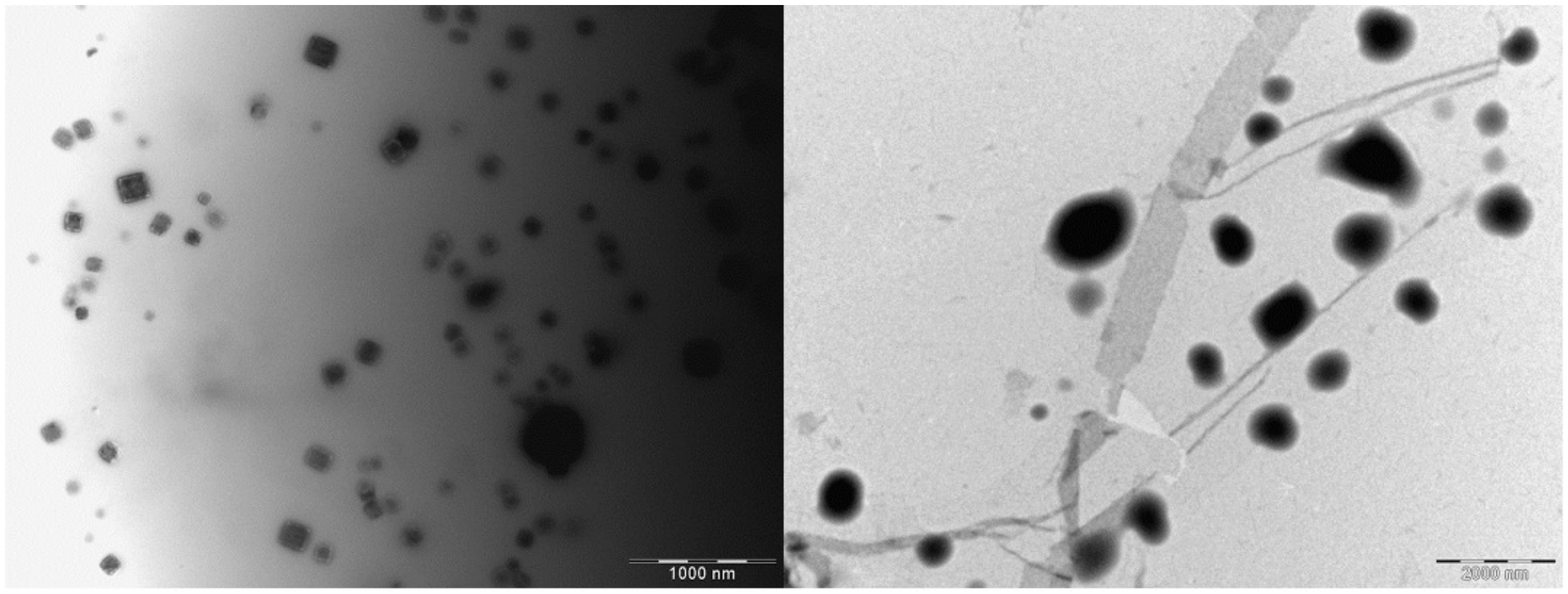}
\centering
\end{figure}  
\end{abstract}

\maketitle

The ability of RADA16 to self-assemble into nanofibers has been studied extensively for use as cell culture scaffolding and drug delivery \citet{ej1, ej2, ej3, fsu3, fsu4}.  It is known that RADA16 conforms into $\beta$-sheets and self-assembles into nano-fibers with widths in the range of 3-10nm in diameter \cite{cormier12,cormier13,yokoi05, arosio12} forming hydrogels when dissolved in water.  Self -assembly produces two distinctive sides:  one side hydrophobic due to alanine, the other hydrophilic due to arginine and aspartic acid \cite{arosio12}.  At least one study has crosslinked a peptide made of a combination of RADA16-Bone morphogenic protein with poly(lactic-co-glycolic acid) via EDC for bone regeneration \cite{pan13}.  Here we study self-crosslinking of the RADA16 peptide via EDC, which could lead to an entirely new range of possible designed peptides with a myriad of functional characteristics.          
\begin{figure} [ht!]
\graphicspath{ {./DisPics/} }
\includegraphics [scale=.15]{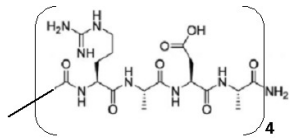}
\centering
\caption {\emph{Acetylated RADA16 with N-terminus \cite{bas13}.}}
\label{semtem}
\end{figure}  

RADA16 studied here is acetylated with an amine N-terminus Ac-[RADA]$_4$-NH$_2$.  The arginine (R) and aspartic acid (D) amino acid residues are positively and negatively charged, respectively.  There are a total of 17 peptide bonds:  15 between RADA amino groups; one at the acetyl end; one at the primary amine.   It contains 17 C=O and 16 N-H bonds in the backbone.  A total of nine sp$^3$ hybridized C bonds stemming from the acetyl end and alanine (A) amino acid subgroups are also present in RADA16.  Side chains in aspartic acid provide a total of four carboxyl groups on the hydrophilic side available for crosslinking by a carbodiimide reaction mechanism.  The N-terminus primary amine is available for crosslinking.  In addition, there are four amines in the arginine guanidinium group that could possibly take part of a crosslinking reaction.  EDC is a zero-length crosslinker which reacts with carboxyl groups to form amine reactive intermediates.  These react with amino groups to form peptide bonds.  An N-substituted urea forms when the intermediate fails to react with the amine \cite{tfs15}.  N-acylurea could also form as a side reaction during crosslinking.  However, the reaction is limited to carboxyls in hydrophobic regions of a protein or polypeptide.  Given that alanine, which forms the hydrophobic region of RADA16 and only contains -CH$_3$, the side reaction was not expected to occur here.         


Figure \ref{semtem}a shows an SEM picture of the resulting product from a reaction between RADA16 hydrogel and EDC prepared as detailed in the Experimental section, both previously dissolved in deionized water.  Nanoparticles of approximately 70-80 nm are readily visible and randomly dispersed throughout the film surface.  To rule out contamination from NaCl or other types of salts, we measured elemental X-ray dispersion with the EDS detector on a 1 $\mu$m x 1 $\mu$m field of view  at four different sample locations.  In addition to elements typical of organic compounds such as carbon, oxygen, nitrogen and hydrogen, EDS measurements showed significant traces of chlorine.  No other elements were found.  We attribute the presence of chlorine to counterions in the RADA16 arginine amino acid residues as well as the hydrochloride from EDC. 
\begin{figure} [ht!]
\graphicspath{ {./DisPics/} }
\includegraphics [scale=.5]{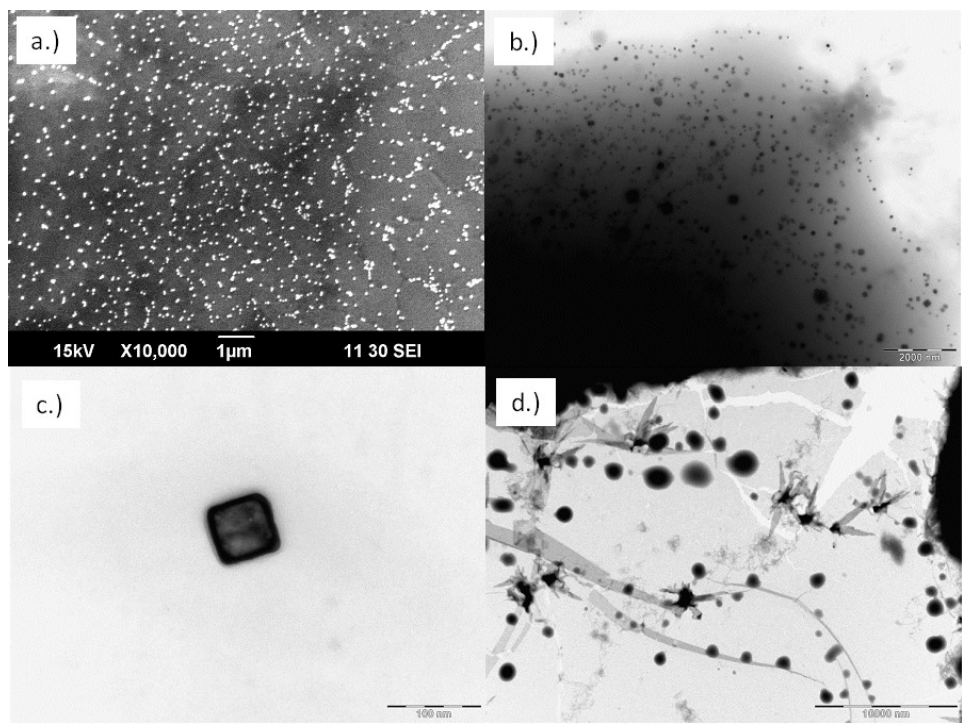}
\centering
\caption {\emph{a.) SEM picture of 2\% w/v RADA16 reacted with 20\% EDC at 10 kX magnification. Monodisperse particles seen throughout sample.  In addition to typical organic elements, EDS measurements showed significant traces of chlorine.  ; b.) Same sample viewed under TEM at 28.7 kX magnification.  Approximately monodisperse orthorhombic crystals visible; c.) TEM close-up view of a $\approx$70 nm nanocrystal at 824 kX magnification.  ; d.) Spherules were also present in sample.  TEM view of spherules at 10.9 kX.  Crosslinked RADA16 nanofibers in process of agglomeration are visible in the middle of picture and lower left corner.  }}
\label{semtem1}
\end{figure}  
Higher magnifications of nanoparticles resulted in pictures that were very fuzzy due to surface charge build-up.  TEM pictures provided better details.  Figure \ref{semtem1}b shows the sample viewed under TEM at 28.7 kX magnification and exhibits a similar nanoparticle monodispersity as seen under SEM.  Under TEM, it is readily apparent that nanoparticles appear to be crystalline in nature and randomly located.  Figure \ref{semtem1}c shows a close-up picture taken with TEM at 824 kX magnification of one of these nanocrystals. This particle appears to have either an  orthorhombic or tetragonal crystal structure.  Studies of additional TEM pictures, led us to believe there is a preponderance of orthorhombic structures with regards to the nanocrystals. Mixed in with the nanocrystals, and somewhat hidden in Figures \ref{semtem1}b and \ref{nta}c, are larger sized spherules.  Figure \ref{semtem1}d presents these spherules, which are bigger in size and in general tend to be $>$ 0.5 $\mu$m.  Interestingly, one could also observe the presence of crosslinked RADA16 nanofibers in process of agglomeration in figure \ref{semtem1}d at the middle and lower left corner of the picture.

\begin{figure}[ht!]
\centering
\graphicspath{ {./DisPics/} }
\includegraphics [scale=0.73]{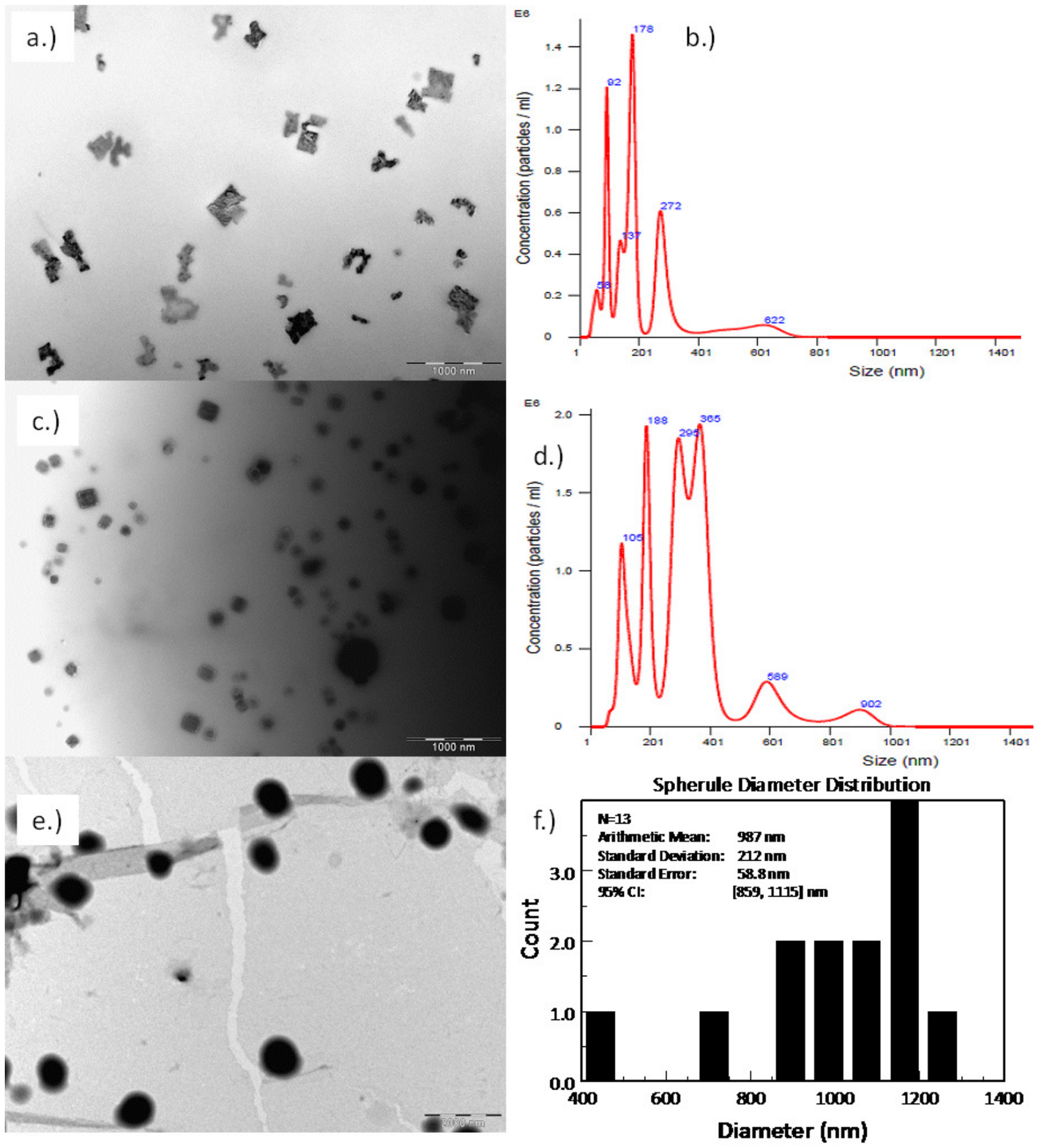}
\caption {\emph{a.) TEM view of plate-like crystals present in 20\% EDC dissolved in deionized water at 78.7 kX magnification.;  b.) NTA measurement of 20\% EDC only, DF=1000, in deionized water measured at 21 $^o$C. ; c.) TEM view of solution made with 2\% RADA16+20\% EDC at 78.7 kX magnification.  Visual inspection shows post-reaction crystals have different morphology than EDC crystals.  Additionally, spherules appear in the mix.  ; d.) NTA measurement of  2\% w/v RADA16+20\% EDC, DF=1000, in deionized water at 25 $^o$C.; e.) TEM view of spherules from different location than figure \ref{semtem1}d at 28.7 kX magnification. ; c.) Spherule size distribution statistics of e.) as measured with the TEM measuring tool. }}
\label{nta}
\end{figure}

To ensure the nanocrystals were not due to unreacted EDC, we measured particle distribution of the reactant using NTA on a sample at DF=1000 in deionized water.  We additionally viewed the same dilution sample under TEM.  Figure \ref{nta}a, TEM picture at 78.7 kX magnification, shows that indeed there are ``plate-like'' square particles or flakes within the EDC solution.  NTA showed particles to be typically in the range of 46 - 300 nm. Less probable were particles of sizes ranging between 500-700 nm. Figure \ref{nta}b presents data for one set of measurements at 21 $^{\rm{o}}$C. A visual comparison of  figure \ref{nta}a with figure \ref{nta}c, also taken with TEM at 78.7 kX magnification and showing reaction product nanocrystals, reveals crystal morphologies are different.  Whereas crystals in EDC are ``plate-like'' flakes at various stages of dissolution, product nanocrystals are solid, well-formed orthorhombic-like structures.  NTA  quantified the size distribution of the mix of nanocrystals and spherules in the product solution.  Figure \ref{nta}d presents data obtained for one set of  measurements from a sample of product solution diluted in deionized water at DF=1000 and measured at 25 $^{\rm{o}}$C.  It shows particles present in the 100-600 nm range within which the majority appear to be nanocrystals.  Larger sizes, $>$ 900 nm, most likely stem from spherules.  Indeed, in a representative area covered primarily with spherules, \ref{nta}e, a manual count of N=13 spherules  yielded an average size D= 987 nm with standard error = 59 nm. The 95\% confidence interval in this region is [859, 1115] nm. Therefore, we attribute the size distribution peaking at 902 nm in Fig. \ref{nta}d to spherules.  Such distribution of sizes did not appear in NTA measurements of EDC.      


\begin{figure}[ht!]
\centering
\graphicspath{ {./DisPics/} }
\includegraphics [scale=0.85]{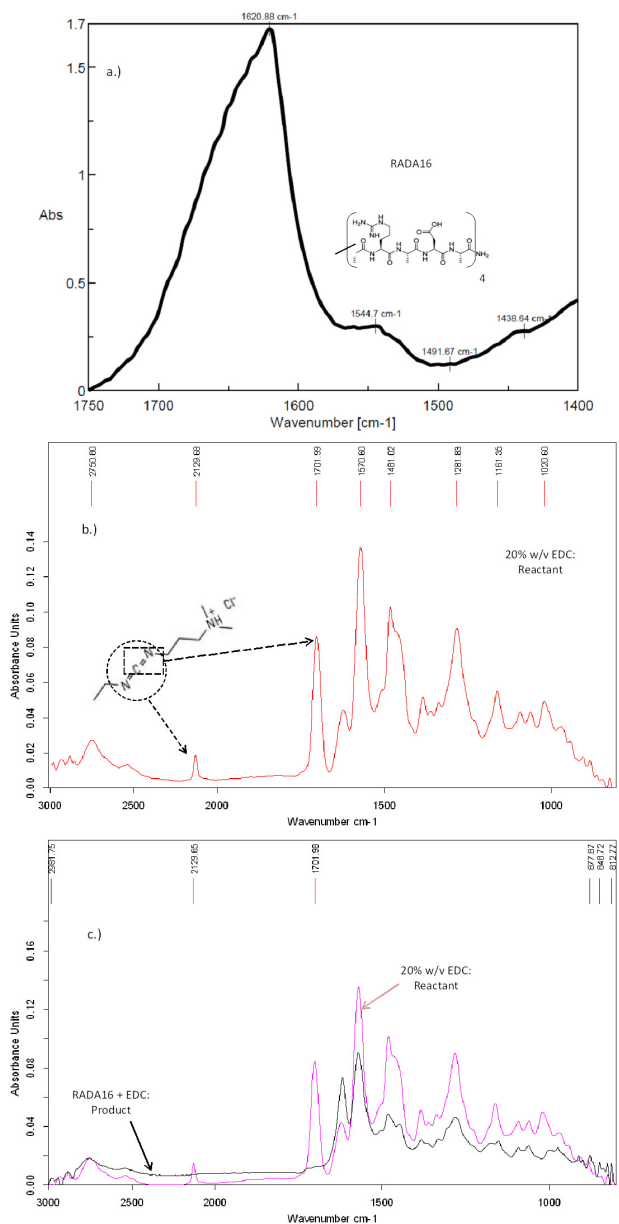}
\caption {\emph{a.) FTIR spectra of 2\% RADA16.  The significant peak at 1636 cm-1 is due to the stable $\beta$-sheets. ; b.) FTIR spectra of 20\% EDC.  N=C=N bonds produce two distinctive peaks at 2130 and 1702 cm-1, respectively.  These disappear after a crosslinking reaction.   ; c.) Overlaid FTIR spectra of unreacted 20\% EDC (magenta) and RADA16+EDC (black) after reaction, diluted in deionized water. }}
\label{ftirl}
\end{figure}

From SEM and TEM pictures, as well as NTA, it was not immediately clear that there was a crosslinking reaction taking place between RADA16 and EDC.  To gather further evidence that the spherules and nanocrystals were not just a result of desegregated RADA16 hydrogel and unreacted EDC, respectively, FTIR measurements were conducted.  

FTIR measurements were obtained for RADA and EDC alone as well as RADA+EDC after reaction.   Figure \ref{ftirl}a is an FTIR plot of RADA16 hydrogel prior to reaction with EDC.  It clearly shows a distinctive $\beta$-sheet peak at 1636 cm-1 \cite{barth07}.  The broad peak at 2116 cm-1 covers the range of alkyne (C$\equiv$C) and nitrile (C$\equiv$N) stretches, which are not thought to be present in RADA16.  Therefore, this broad peak is unknown as of this writing.  Figure \ref{ftirl}b shows FTIR data for EDC prior to reaction with EDC.  Of particular importance are the peaks at 2130 and 1702 cm-1 as these distinctive peaks for EDC disappear after the crosslinking reaction with RADA16.  The peak at 2130 cm-1 is attributed to the N=C=N bonds of EDC \cite{mit60}.  We attribute the peak at 1702 cm-1 to stretching of the cumulated C=N bonds since C is an sp hybridized carbon.  It is expected that these bonds would no longer be present after reaction of the primary amine with the unstable intermediate o-acylisourea.     That is in fact what we found.  Figure \ref{ftirl}c shows overlaid plots of EDC reactant (magenta) with the RADA16+EDC product (black).  Peaks at 2130 and 1702 cm-1 are conspicuously absent, confirming that a crosslinking reaction indeed took place. It is also clearly obvious that the $\beta$-sheet peak at  1636 cm-1 present in RADA16 no longer appears after crosslinking.  Evidently, the stable $\beta$-sheet structure of RADA16 has been disrupted by the crosslinking mechanism. The two most prominent peaks of the RADA16+EDC product (black curve) appear at 1619 and 1571 cm-1.  Both peaks could be attributed to different modes of N-H bending vibrations of primary amine groups such as those present in urea. Given the relative signal strength of the peaks, it is also possible that they are produced by bending vibrations of amide groups. Crosslinking confirmed, we would expect the peaks to be produced by the presence of both amide bonds from crosslinked peptides and amine bonds from the urea by-product.  The peak at 1571 cm-1 could also be produced by a nitro group (-NO$_2$) asymmetric stretch, though we believe this type of bond is less likely to occur in the current crosslinking environment.  The next highest peaks of the RADA16+EDC product curve appear at 1481 and 1281 cm-1.    Bending and rocking deformations of alkane groups (CH$_2$; CH$_3$) are attributed to 1481 cm-1.     It seems likely that the C-N stretch of an amine group causes the peak at 1281 cm-1.  The low energy peaks at 878, 849 and 813 cm-1 which appear on the product curve, but not on the EDC curve could be due to C-Cl bond stretching.  However, that needs to be studied in more detail and we will not mention them further.  On the higher energy side of the product spectrum we attribute the peak at 2982 cm-1 to sp$^3$ hybridized C-H bonds signifying the presence of acetyl groups.  It seems the  2754 cm-1 peak on the RADA16/EDC product curve could be produced by the  O-H stretch of  regenerated carboxylic acids that did not react with a primary amine.  It is not likely that the peak is produced by the C-H stretch of aldehydes.  

FTIR data, thus, lends support to the existence of a proposed crosslinking reaction of RADA16 activated by EDC.  It is likely that crosslinking proceeds through EDC activation of the carboxyl groups  present in the aspartic acid amino acid residues.  The unstable, amine-reactive O-acylisourea intermediate that results from activation of the arboxyl groups then reacts with available primary amines.  Primary amines available for reaction either come from the N-terminus or the guanidinium group of the arginine subgroup.  While the guanidinium cation is highly stable in an aqueous solution, reactions stemming from a combination of both the N-terminus and possibly guanidinium groups cannot be ruled out. 

It should be pointed out that Powder XRD analysis of the product yielded inconclusive results. It was not possible to obtain significant readings with the amount of raw material at hand.  Currently available TEM is not equipped with XRD capability to further analyze nanocrystalline structures.


RADA16 was obtained from 3D Matrix as a lyophilized powder that was prepared by exchanging TFA for HCl \cite{bas13} so that the arginine had a chlorine counterion and the aspartic acid was protonated.   It was reconstituted in deionized water at a nominal 2.0\% (w/v) to give a solution with pH $\approx$ 2-3.  

EDC was obtained from TCI America (USA) as a hydrochloride with a MW=191.70 g/mol and of 98.0\% purity.  It was dissolved in deionized water to obtain a nominal 20 \% (w/v) solution with pH~7.68 as measured with a Sensorex polymer electrode.     

Procedure for reacting RADA16 with EDC was the following.  To 100$\mu$L of 2\% (w/v) RADA16 gel we added  50$\mu$L of 20\% (w/v) EDC.  The mixture was shaken vigorously for approximately 5 minutes on a Vortex Genie mixer at setting 7 then placed in a lab bench Fisher-Scientific centrifuge for two minutes . To improve mixing, we let the mixture sit overnight for approximately 24 hrs.  The resulting aqueous solution had a pH=3.53.  All reactions were carried out at 22 $^{\rm{o}}$C.   

In preparation for viewing the sample under SEM, 250 $\mu$L of 70\%  (w/v) ethanol was added to reactant mixture to both dissolve any unreacted polymer and aid in evaporation of the solution.  Approximately 100 $\mu$L of the solution was placed on a coverglass that was cleaned by immersion in ethanol and sonicated for 10 minutes.  The product solution on the coverglass was then evaporated for about 6 minutes on top of a hotplate set at 90 $^{\rm{o}}$C.  To view under SEM, an approximately 10 nm layer of Au-Pd was deposited on top of the dried RADA16/EDC film with a Denton sputtering system.  

Preparation of samples for viewing under TEM required nominal dilution factors of DF=1000.  Samples were vacuum dried at 45 $^{\rm{o}}$C and negatively died.  

Nanoparticle Tracking Analysis equipment required volumes in the range of 0.8-1 mL.  We used dilution factors of DF=1000 in deionized water to study the distribution of particle sizes in our sample.  

FTIR studies were conducted at room temperature, 22 $^{\rm{o}}$C, using the same dilution factor.

A JEOL JSM-63900LV SEM equipped with an energy-dispersive X-ray spectroscopy (EDS) detector from Oxford Instruments was used to obtain SEM pictures and material composition data.  RADA16 FTIR spectra were obtained at 4 cm-1 resolution on a Jasco FT/IR 4100 with a multi-reflection Attenuated Total Reflectance (ATR) accessory equipped with a ZnSe crystal.  FTIR spectra for EDC and RADA16+EDC product were measured on a Bruker Vertex 70 spectrometer with a single pass ATR accessory.  Nanoparticle Tracking Analysis (NTA) was performed on a Malvern Instruments Nanosight LM10 with capability of tracking particles in the size range of 10 - 2000 nm.  TEM data was obtained in collaboration with the Microscopy Core Facility.

We have provided evidence that crosslinking in RADA16 is activated by EDC.  It is likely that crosslinking proceeds through EDC activation of the carboxyl groups  present in the aspartic acid amino acid residues reacting with primary amines either from the N-terminus and possibly the guanidinium group of the argininine subgroup.  The reaction produces nanocrystals and micron-sized spherules.  Further studies are required to understand the mechanisms leading to crosslinking as well as formation of nanocrystals and spherules.


 This work has been supported in part by the Microscopy Core Facility in the Department of Integrative Biology at the University of South Florida.  We also like to thank Dr. Haynie for very useful comments.

\bibliography{rada_edc}

\end{document}